\documentclass[12pt, oneside, letterpaper]{article}

\usepackage{setspace,graphicx,epstopdf,amsfonts,amssymb,amsthm,versionPO}
\usepackage{marginnote,datetime,enumitem,subfigure,rotating,fancyvrb}
\usepackage{commath}
\usepackage{amsmath,bm}
\usepackage{algorithm}
\usepackage{algpseudocode}
\usepackage[longnamesfirst]{natbib}
\usdate

\excludeversion{notes}		% Include notes?
\includeversion{links}          % Turn hyperlinks on?

\iflinks{}{\hypersetup{draft=true}}

\usepackage[small, md, sc]{titlesec}
\usepackage[tight,nice]{units}
\RequirePackage{verbatim}

\RequirePackage{color}
\definecolor{Crimson}{rgb}{0.6471, 0.1098, 0.1882}

\ifnotes{%
\usepackage[margin=1in,paperwidth=10in,right=2.5in]{geometry}%
\usepackage[textwidth=1.4in,shadow,colorinlistoftodos]{todonotes}%
}{%
\usepackage[margin=1in]{geometry}%
\usepackage[disable]{todonotes}%
}

\makeatletter\let\chapter\@undefined\makeatother % Undefine \chapter for todonotes

\usepackage{indentfirst} % Indent first sentence of a new section.
\usepackage{endnotes}    % Use endnotes instead of footnotes
\usepackage{jf}          % JF-specific formatting of sections, etc.
\usepackage[labelfont=bf,labelsep=period]{caption}   % Format figure captions
\usepackage{amsmath}
\graphicspath{ {images/} }

\usepackage{geometry}
 \usepackage{chngcntr}
\counterwithin{figure}{section}
\numberwithin{equation}{section}

\usepackage{float}
\usepackage{titlesec}

\titleformat{\paragraph}
{\normalfont\normalsize\bmseries}{\theparagraph}{1em}{}
\titlespacing*{\paragraph}
{0pt}{3.25ex plus 1ex minus .2ex}{1.5ex plus .2ex}

 \usepackage[subnum]{cases}

\usepackage[english]{babel}
\usepackage[autostyle]{csquotes}

\usepackage{booktabs}
\usepackage[bottom]{footmisc}

\begin{document}

\setlist{noitemsep}  % Reduce space between list items (itemize, enumerate, etc.)
\onehalfspacing      % Use 1.5 spacing
% Use endnotes instead of footnotes - redefine \footnote command
\renewcommand{\footnote}{\endnote}  % Endnotes instead of footnotes

\author{\large \textit{Ali Habibnia}\thanks{\rm Department of Economics, Virginia Tech. Email: habibnia@vt.edu} \and {\large \textit{Esfandiar Maasoumi}}\thanks{\rm Department of Economics,  Emory University.}}

\title {\textcolor{Crimson}{\Large \textit {Forecasting in Big Data Environments:} {\normalsize \textit an Adaptable and Automated Shrinkage Estimation of Neural Networks (AAShNet)}}}

\date{}    % No date for final submission

% Create title page with no page number

\maketitle
\thispagestyle{empty}

\bigskip

\centerline{ABSTRACT}

\begin{onehalfspacing}  % Double-space the abstract and don't indent it
\noindent This paper considers improved forecasting in possibly nonlinear dynamic settings, with high-dimension  predictors (“big data” environments). To overcome the curse of dimensionality and manage data and model complexity, we examine shrinkage estimation of a back-propagation algorithm of a neural net with skip-layer connections. We expressly include both linear and nonlinear components. This is a high-dimensional learning approach including both sparsity $L_1$ and smoothness $L_2$ penalties, allowing high-dimensionality and nonlinearity to be accommodated in one step. This approach selects significant predictors as well as the topology of the neural network. 
%We estimate shrinkage hyper-parameters by minimizing a validation error in the training of neural networks. 
We estimate optimal values of shrinkage hyperparameters by incorporating a gradient-based optimization technique resulting in robust predictions with improved reproducibility. The latter has been an issue in some approaches. 
This is statistically interpretable and unravels some network structure, commonly left to a black box. An additional advantage is that the nonlinear part tends to get pruned if the underlying process is linear. In an application to forecasting equity returns, the proposed approach captures nonlinear dynamics between equities to enhance forecast performance. It offers an appreciable improvement over current univariate and multivariate models by RMSE and actual portfolio performance.

\end{onehalfspacing}

\clearpage

\begin{doublespace}
\section{Introduction}
An important step in designing modern predictive models is to cope with high-dimensional data, presenting large numbers of (cor)related variables and complex properties. ``Big data" is both an increase in the number of samples collected over time, and an increase in the number of potential explanatory variables and predictors. When dimension grows, the specificities of high-dimensional spaces and data must then be taken into account in the design of predictive models. While this is valid in general, its importance is heightened when using nonlinear tools such as artificial neural networks. Most nonlinear models involve more parameters than the dimension of the data space which may result in a lack of identifiability, lead to instability, and overfitting  (\cite{huber2011};\cite{fromstatisticstoneuralnetworkstheoryandpatternrecognitionapplications1994}; \cite{moody1991}). Selection of significant predictors, and model complexity are the key tasks of designing accurate predictive models in data-rich environments.

Feature extraction and feature selection are broadly the two main approaches to dimensionality reduction. Extraction transforms the original features into a lower dimensional space preserving all its fundamentals. Feature selection methods select a small subset of the original features without a transformation. Extraction methods include principal component analysis - \cite{pearson1901}; \cite{eckart1936}; factor analysis - \cite{spearman1904}; canonical correlations analysis - \cite{hotelling1936}, and several others\footnotemark\footnotetext{\cite{cunningham2015} surveyed the literature on linear dimensionality reduction in their work.}.). Feature selection is accomplished by such methods as Ridge - \cite{hoerl1970}; LASSO - \cite{tibshirani1996} and Elastic Net - \cite{zou2005}).  

In this work, our main focus is on feature selection techniques. We apply shrinkage approaches (usually referred to as regularization in machine learning literature). We embed feature selection in the backpropagation algorithm as part of its overall operation. 
Accordingly, we extend our loss function to include $L_1$ norm for the weights of the dense network, and $L_2$ norm for the weights in the skip-layer. The dense network corresponds to a multilayer neural network, whereas the skip-layer denotes the direct connection from each of the input variables to each of the output variables, which is similar to a linear regression model. 
%To be more precise, we penalize the neural network loss function with the $L_1$ norm for weights in the hidden layer and $L_2$ norm for weights in the skip-layer. The latter represents direct connections from each of the input variables to each of the output variables. This part of the model is equivalent to a linear regression.

Shrinkage is an implicitly embedded feature selection. It is an example of model selection since only a subset of variables contributes to the final predictor. It has frequently been observed that $L_1$ shrinkage produces many zero parameters, leading to some features being dropped and a sparse model. Only those parameters whose impact on the empirical risk is considerable appear in the fitted model \cite{ng2004}. Shrinkage is a proper means of controlling complexity in the nonlinear component. From an optimization point of view we have a neural network learned/estimated by LASSO. This prevents hidden units from getting stuck near zero and/or exploding weights. 

Simultaneously, we employ the $L_2$ shrinkage on the skip-layer connections (linear part of the model), in order to penalize groups of parameters, and encourage the sum of the squares of the parameters to be small. Therefore we will not drop specific features from linear component, making it possible to interpret the marginal impact of predictors on the target variable. It is worth mentioning that the linear part of the model can be interpreted as a Ridge regression. 

There are other benefits to shrinkage/regularization. Empirically, penalizing the magnitude of network parameters is also a way to reduce overfitting and to increase prediction accuracy \cite{ng2004}. This is especially true in the state-of-art models, such as deep learning models with large number of parameters. Our proposed algorithm combines the neural network’s advantage of describing the nonlinear process with the superior accuracy of feature selection that is provided by a penalized loss function that combines $L_1$ and $L_2$ norms.

Many studies have suggested neural networks as a promising alternative to linear regression models. Empirical evidence on out-of-sample forecasting performance is, however, mixed. It is challenging to determine linear or nonlinear components.
Linearity tests do often suggest that real world series are rarely purely linear or nonlinear. 

We consider the possibility that the series $(y_t)$ contain both a linear component, $(\mathcal{L}_t)$, and a nonlinear component $(\mathcal{N}_t)$.

\begin{equation} \label{eq1}
y_t= \mathcal{L}_t + \mathcal{N}_t
\end{equation}

Neural network alone is not best suited to handle both linear and nonlinear components, especially when the linear component is superior to the nonlinear component. 

Two different approaches to model and forecast series with both linear and nonlinear patterns are available. The first approach is a two step methodology to combine linear time series models and neural network models. In this approach, the first step residuals are obtained from the fitted linear model $\hat{e_t}= y_t - \hat{\mathcal{L}_t}$. In the second step a nonlinear model (e.g., GARCH, neural nets) is trained on the residuals of the first step. In principle, this ``hybrid'' two step approach can provide superior predictions when both the linear and neural network model are well specified. In practice, however, two types of model specification errors are introduced without an ability to assess their mutual impact.

The alternative approach that we are proposing in this paper models both linear and nonlinear components adoptively. It is based on a neural network with skip-layer connections including both linear and nonlinear structures. 

The rest of the paper is organized as follows. Section II provides the basic framework of the proposed model. In Section III we investigate proper estimation of shrinkage hyperparameters and introduce gradient-based techniques based on reverse-mode automatic differentiation (RMAD) to accomplish this. Section IV presents an application to US financial returns. Section V contains some concluding remarks.

\section{The Model} \label{sec:Model}

%In this study, we focus on feedforward neural networks with only one hidden layer. And to show that the neural network models can be seen as a generalisation of linear models, we allow for direct connections from the input variables to the output layer and we assume that the output transfer function is linear\footnotemark \footnotetext{Using linear function for the output unit activation function (in conjunction with nonlinear activations  amongst the hidden units) allows the network to perform a powerful form of nonlinear regression. So, the network can predict continuous target values using a linear combination of signals that arise from one layer of nonlinear transformations of the input.}, then the model becomes

\begin{figure}[htp]
\centering
\includegraphics[width=14cm]{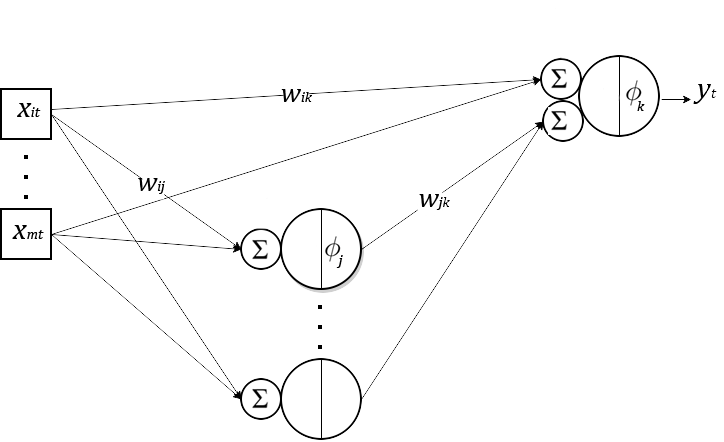}
\caption{A single-hidden-layer neural network with skip-layer connections}
\label{fig:lion}
\end{figure}

In this study, we examine a feedforward neural network with one hidden layer, known as a dense network. Neural network models can be seen as generalizations of linear models, when one allows direct connections from the input variables to the output layer with a linear transfer function\footnotemark \footnotetext{Using linear function for the output unit activation function (in conjunction with nonlinear activations amongst the hidden units) allows the network to perform a powerful form of nonlinear regression. So, the network can predict continuous target values using a linear combination of signals that arise from one layer of nonlinear transformations of the input.}, that we refer to as the skip-layer. The model is expressed as

\begin{equation} \label{NN}
y_t= \Phi({\bm x};{\bm w}) =  \sum_{i\rightarrow k} x_{it}w_{ik} +\sum_{j\rightarrow k} \phi_j  \bigg(\sum_{i\rightarrow j} x_{it} w_{ij} \bigg)w_{jk}+\varepsilon_t,
\end{equation}

where $\Phi$ describes the network by a vector function. We associate subscript $i$ with the input layer, subscript $j$ with the hidden layer, and subscript $k$ with the output layer. $x_{it} = (x_{1t},x_{2t},...,x_{mn})$ is the value of the $i$th input node, which can be a constant input representing biases, a matrix of lagged values of $y_t$ and some exogenous variables. $\phi_j(.)$ and $J$ are activation functions and number of neurons used at the hidden layer. A single-hidden-layer neural network with skip-layer connections is shown in Figure \ref{fig:lion}. A network with only one hidden layer and skip-layer connections has three sets of weights:
those for direct connections between the inputs and the output ($w_{ik}$), those connecting the inputs to the hidden layer ($w_{ij}$), and those connecting the output of the hidden layer to the final output layer($w_{jk}$).  

First term in Eq.\eqref{NN} represents a linear regression term. The second term, denoting the dense network of the two layers, hidden and output, is usually referred to as a multi-layer perceptron in the literature. It has been shown to be able to perform well with nonlinear complex data. A greater capacity of the dense network, compared to the skip-layer, is realized by stacking two layers, enabling it to model more complex data. A differentiable nonlinear activation function $\phi$ is used in the hidden units. $\varepsilon_t$ is a random disturbance term which captures all other factors influencing $y$ than the $x$. A linear component term moves the model in the linear direction. This aids statistical interpretation and unravels the structure behind the network, otherwise left to a black box. This simultaneous approach has the advantage, when we apply shrinkage techniques to estimate network parameters for an essentially linear process, of pruning the hidden neurons. 

Estimation of network elementary parameters based on prediction error minimisation is known as training/learning. The most common cost/risk function is the mean squared prediction error (MSE), $E = \frac{1}{n}\sum_{t=1}^n (y_t - \hat{y}_t)^2$. Given target values $y_t$ and network estimated outputs $\hat{y}_t$ error functions are obtained for each parameter set, followed by tuning of the parameters.

The error surface becomes increasingly complicated with the number of input variables and network parameters. It is common to employ the conventional feed-forward neural network, trained with the popular and revolutionary gradient-descent-type algorithm known as backpropagation. The backpropagation algorithm was first introduced by \cite{bryson1979} and popularized in the field of artificial neural network research by \cite{werbos1988} and \cite{rumelhart1986}. Error function's sensitivity to network parameters is assessed via Gradient Descent optimization. Gradient is normally defined as the first order derivative of the error function with respect to each of the model parameters. Working out the gradients can be performed in a completely mechanical way known as Automatic Differentiation \cite{baydin2017automatic}. AD employs the Jacobian matrix of gradients for each parameter $w_i$ to identify directions that decrease the height of the error surface (see Appendix). In fact backpropagation is only a specific case of reverse-mode AD that is applied to an objective function errors as functions of model parameters.
%Iterative calculation of gradients and updating the parameter will reach a local minimum in error landscape. 
The weight adjustment is given by

% \begin{equation} \label{eq:delta rule}
% w^{new} = w^{old} - \eta~ \frac{\partial E(\bm{w})}{\partial w}
% \end{equation}

\begin{equation} \label{eq:delta_rule}
w^{new} = w^{old} - \eta~ \frac{\partial E(\bm{w})}{\partial w}
\end{equation}

Where the constant $\eta$ is the learning rate (step size) for updating elementary parameters, its value falls between zero and one. By iteratively repeating this mechanism, the network can be trained in a way that converges to the optima. The set of new elementary parameters are repeatedly presented to the network until the error value is minimized. Around the optimum point, all the elements of the gradient would be very small, leading to tiny changes in new parameters. 

%Where the constant $\eta$ is the learning rate (step size) and its value falls between zero and one.  The direction of search in weight space for the new value of the weights is elected by $\frac{\partial E(\bm{w})}{\partial w}$, that shows the sensitivity of the error function to the weights. By repeating iteratively the steps network can be trained in a way that converges to optima. The set of new weights are repeatedly presented to the network until the error value is minimised. Around the optimum point, all the elements of the gradient would be very small, which leads to tiny changes in new weights. 

%Implementing $L_1$ and $L_2$ regularization in a backpropagation algorithm of a neural network is relatively easy. In particular, method error returns the  total error plus penalties or constraints and the objective of learning  is minimization of the regularized loss function. If weight values are included in the total error term that’s being minimized, then smaller weight values will generate smaller error values. Therefore, network parameters are the solutions to the following optimization problem

We add the $L_1$ and $L_2$ penalties in training our modelto the loss function $\widetilde{E}(.)$, the original MSE. The following optimization problem is used for training:

% \begin{equation} 
% {\bm w}^* = \underset{{\bm w}} {\DeclareMathOperator{\argmin}{\arg\!\min}}~ E(\bm{w}) ~+~ \lambda~\Omega(\bm {w})
% \end{equation}

\begin{equation} 
\begin{aligned}
{\bm w}^* &= \underset{\bm w}{\arg\!\min}~ \widetilde{E}(\bm{w}|\lambda, X) = \underset{\bm w}{\arg\!\min}~ E(\bm{w}|\lambda, X) ~+~ ~\Omega(\bm {w,\lambda})\\
%&= \underset{{\bm w}} {\DeclareMathOperator{\argmin}{\arg\!\min}}~ E(\bm{w}|\lambda, X) ~+~ \frac{\lambda_2}{2}\sum_{i\rightarrow k} w^2_{ik}~+~\lambda_1(\sum_{i\rightarrow j} |w_{ij}|+\sum_{j\rightarrow k}|w_{jk}|)
\end{aligned}
\end{equation}

%where the regularization term $\Omega(\bm {w})$ is multiplied by a shrinkage/regularization hyperparameter $\lambda$. Assuming a fixed $\lambda$, to learn network parameters ${\bm w}^*$ using backpropagation algorithm, we just need to add derivative of penalty term to the gradient and iteratively update parameters\footnotemark \footnotetext{It is worth noting that the regularization term does not include the biases.}.   

where the regularization term $\Omega(\bm {w, \lambda})$ is a combination of the L1 norm and the L2 norm of the parameter vector. $\lambda$ sets the impact of shrinkage on the loss, with larger values resulting in more penalization. Using the regularized objective causes the training procedure to be inclined to smaller parameter values; unless larger parameters considerably improve the original error value (MSE). Assuming a fixed $\lambda$, to learn $w^*$, we only need to include the derivative of $\Omega(\bm {w, \lambda})$ in our derivatives:

\begin{equation} \label{eq:delta_rule2}
\left\{
  \begin{array}{lr}
    \Delta = \frac{\partial E({\bm w})}{\partial w}  +  \frac{\partial \Omega({\bm w, \lambda})}{\partial w}  \\
  w^{new} =  w^{old} ~ - ~ \eta \Delta
  \end{array}
\right.
\end{equation}

Where $\Delta$ is the gradient of the regularized loss function. $\lambda > 0$ is proportional to complexity of the model but is not a parameter that appears in the model. It is a hyperparameter. In the next section, we explain the impact of hyperparameters and elaborate on our procedure for tuning them.

%Then the parameter updating rules become 
We employ $L_1$ and $L_2$ shrinkage on the parameters of the dense network and skip-layer, respectively; as is depicted by following optimization problem:
\begin{equation} 
{\bm w}^* = \underset{\bm w}{\arg\!\min}~ E(\bm{w}|\lambda, X) ~+~ \frac{\lambda_2}{2}\sum_{i\rightarrow k} w^2_{ik}~+~\lambda_1(\sum_{i\rightarrow j} |w_{ij}|+\sum_{j\rightarrow k}|w_{jk}|)
\end{equation}

which can be realized by iteratively adjusting the parameters using the updating rules below

\begin{equation} \label{eq:delta_rule3}
\left\{
  \begin{array}{lr}
    w_{ik}^{new} = w_{ik}^{old} - \eta~ (\frac{\partial E({w|\lambda, X})}{\partial w_{ik}}
     + \lambda_2w_{ik}^{old})\\
    
    w_{ij}^{new} = w_{ij}^{old} - \eta~ (\frac{\partial E({w|\lambda, X})}{\partial w_{ij}}
     + \lambda_1sgn(w_{ij}^{old}))\\
    
    w_{jk}^{new} = w_{jk}^{old} - \eta~ (\frac{\partial E({w|\lambda, X})}{\partial w_{jk}}
     + \lambda_1sgn(w_{jk}^{old}))\\
  \end{array}
\right.
\end{equation}

%Where $\Delta$ is the gradient of regularized loss function and $\lambda >0$ is known as the regularization hyperparameter. In many practical applications, the simplest method to set $\lambda$ is to train the neural network with a number of different $\lambda $ values, and then choose the model having the smallest validation error. A more attractive approach is to use an optimization algorithm to adapt the model hyperparameters $\lambda$ automatically. In the next section we will explain how to estimate a set of $\lambda$ for each network weight in order to minimize a validation error during training of neural networks. 

Where $\lambda_1$ and $\lambda_2$ are non-negative values known as shrinkage hyperparameters. $L_1$ sparsity norm and $L_2$ smoothing norm are two closely related regularizers that can be used to impose a penalty on the complexity of the model that is to be learned. Shrinkage estimation of the model can be seen as an implementation of Occam's razor, introducing a controllable trade-off between fitting data and model complexity, enabling us to have models of less complexity with adequate generalization capability. Regularization in neural networks limits the magnitude of network parameters by adding a penalty for weights to the model error function. In this study, $L_2$ shrinkage penalizes parameters in skip-layer connections by adding sum of their squared values to the error term. $L_1$ shrinkage penalizes parameters in the dense network to encourage the topology of the learned network to be sparse. The relative importance of the compromise between finding small weights and minimizing the original risk function depends on the size of $\lambda$.

To use $L_2$ shrinkage, we add a $\lambda_2 w$ term to the gradient as the derivative of $w^2$ is $2w$. $L_2$ shrinkage works with all forms of learning algorithms, but does not provide implicit feature selection. The derivative of the absolute value of $w$ is $w/|w|$, however $L_1$ norm is not differentiable at zero and hence poses a problem for gradient-based methods.

The problem can be solved using the exact gradient, which is discontinuous at zero. We can also solve the problem by the smooth approximation approach which will allow us to use gradient descent. To smooth out the $L_1$ norm using an approximation, we use $\sqrt{w^2 +\epsilon}$ in place of $|w|$ , where $\epsilon$ is a smoothing parameter which can also be interpreted as a sort of sparsity parameter. When $\epsilon$ is large compared to $w$, the expression $w + \epsilon$ is dominated by $\epsilon$ and taking the squared root yields approximately $\sqrt{\epsilon}$. \cite{lee2006}

% This is exactly the same as the usual gradient descent learning rule, except we rescale the weights that makes them smaller. Intuitively, the effect of penalty terms is to lead the network to learn small weights, all other things being equal. Large weights will only be allowed if they considerably improve the first part of the cost function. The relative importance of the compromise between finding small weights and minimizing the original loss function depends on the size of $\lambda$. When $\lambda$ is small we prefer to minimize the original loss function, but when it is large we prefer small weights. There are different ways to tune hypeparameters such as $\lambda$. We explain later how to estimate $\lambda$ instead of setting that manually using grid search or cross-validation.

\section{Gradient-based Hyperparameter Optimization}

The major drawback of shrinkage is that it introduces additional hyperparameters. In practice we have two set of parameters: model elementary parameters (network weights and biases), and learning algorithm hyperparameters (magnitude of $L_1$ and $L_2$ penalties, and learning rate). We would ideally like to determine these hyperparameters to get optimal generalization\footnotemark \footnotetext{Generalization means building a model on one set of training data and hope that it makes effective predictions on a different set of test data.}. As opposed to elementary parameters, these hyperparamters cannot be directly trained by the data. Whereas the elementary parameters specify how to transform the input data into the desired output, the hyperparameters define how our model and algorithm are actually structured.

The performance and robustness of neural networks relies to a large extent on hyperparameters. Tuning these hyperparameters not only makes the investigation of methods difficult, but also hinders reproducibility (\cite{bergstra2011algorithms}). Transparent tuning of hyperparameters can be part of an Hyperparameter Optimization (HPO), as an outer loop in training procedures. 

%% RANDOM AND GRID SEARCH
The de-facto na\"ive approach of searching through combinations of potential values of hypergradients and choosing the one that performed the best (a.k.a. grid search) is very time-consuming and becomes quickly infeasible as the dimension of hyperparameter space grows. In many practical applications manually searching the space of hyperparameter settings is tedious and tends to lead to unsatisfactory outcomes. \cite{bergstra2012} show empirically and theoretically that random search more efficient than grid search. Statistical techniques such as cross-validation \cite{wahba1990}, bootstrapping \cite{efron1994}, and Bayesian methods \cite{mackay1992} can also assist in determining hyperparameters. 

%% CROSS VALIDATION
HPO must be guided by some performance metric, typically measured by cross-validation (CV) on the training set, or evaluation on a held-out validation set. The rationale behind CV is to split the data into the training samples used for learning the algorithm, and the validation samples (one or several folds) for estimating the risk of each algorithm and for evaluation of its performance. CV consists of averaging several hold-out estimators (folds) of the risk corresponding to different splits of the data, and selecting the algorithm with the smallest estimated risk. Within each fold, hyperparameters are fixed and we only estimate model elementary parameters. The validation samples play the role of new unseen data as long as the data are i.i.d.\footnotemark \footnotetext{This assumption can be relaxed. see: \cite{chu1991}.} For a general description of the CV see \cite{geisser1975}, and \cite{arlot2010} for a comprehensive review on cross-validation procedures and their applications in different algorithms and frameworks. Several studies such as \cite{rivals1999} show cases in which CV performance is less than satisfactory.

%% GRADIENT-FREE HPO
Recently, automated approaches for estimation of hyperparameters have been proposed which can provide substantial improvements and transparency. Although one may also ``hyperparameterize'' certain discrete choices in design of the model (e.g. number of hidden units), we focus only on the continuous hyperparameters in this work. There are a number of gradient-free automated optimization methods (\cite{hutter2011}; \cite{bergstra2011}; \cite{bergstra2013}; \cite{snoek2012}), all of which rely on multiple complete training runs with varied fixed hyperparameters. Hyperparameters are chosen to optimize the validation loss after complete training of the model parameters.

%% INTRO TO GRADIENT-BASED HPO
Gradient-based HPO approaches, proposed by \cite{larsen1996} and \cite{andersen1997}, emerged in the 1990s. 
% They formulate an iterative gradient descent scheme for adapting the hyperparameters by minimizing the validation error calculated from a single validation set. They treat hyperparameters similar to elementary parameters during training and simultaneously update both sets of parameters. Following this scheme, we can estimate a single regularization parameter or separate regularization parameter for each individual weight in the network.
We can distinguish two main approaches of gradient-based optimization: Implicit differentiation and iterative differentiation. 

%% IMPLICIT DIFFERENTIATION
Implicit differentiation, first proposed by \cite{larsen1996}, computes the derivative of the cost $L_{valid}$ with respect to $\lambda$ based on the observation that, under some regularity conditions, the implicit function theorem can be applied in order to calculate the gradients of the loss function. In particular, the cost function is assumed to smooth and converge to local minima. The inner optimization $w(\lambda) \in \textup{argmin}_{w}L_{train}$ can be characterized by the implicit equation $\nabla_w L_{train} = 0$. \cite{bengio2000gradient} derived the gradients for unconstrained cost function and applied the algorithm to $L2$ shrinkage for linear regression. The method has also been used to find kernel parameters of Support Vector Machines \cite{keerthi2007efficient}. \cite{pedregosa2016hyperparameter} proposes \textit{HOAG} which uses inexact gradients, allowing the gradient with respect to hyperparameters to be computed approximately. 

%% ITERATIVE DIFFERENTIATION
In iterative differentiation, first proposed by \cite{domke2012generic}, the gradient for hyperparameters are calculated by differentiating each iteration of the inner optimization loop and using the chain rule to aggregate the results. However, the problem with this reverse-mode approach is that one must retain the entire history of elementary parameter updates, making a na\"ive implementation impractical due to memory constraints. Reverse-mode differentiation requires intermediate variables to be maintained in the memory for the reverse pass and evaluation of validation loss needs hundreds or thousands of inner optimization iterations. \cite{maclaurin2015} later extended this for setting of stochastic gradient descent via reverse mode automatic differentiation of validation loss.The burden of storing the entire training trajectory $w_1,\cdots,w_T$ is avoided by an algorithm that exactly reverses SGD with momentum to compute gradients with respect to all training parameters, only using a relatively small memory footprint, making a solution feasible for large-scale big data machine learning problems. 

%Automatic differentiation can provide us with a mechanical way of calculating gradients with respect to hyperparameters, often referred to as hypergradients. 
We defined the updating rule for elementary parameters as $w_{t+1} = w_t - \eta \nabla L_{train}$ where $L_{train} = \widetilde{E}(w_t|\lambda, X_{train})$ is the regularized loss value on train data. To calculate hypergradients we rely on the unregularized loss function, that is $L_{valid} = E(w_t|\lambda, X_{valid})$, as the actual generalization performance of the model, on unseen data points, does not directly depend on regularizers; otherwise the model with no regularization would be always selected:

\begin{equation}
\begin{aligned} 
\lambda^{*} \; &= \textup{argmin}_{\lambda}L_{valid}\\
\textup{s.t.\;\;} w(\lambda) &\in \textup{argmin}_{w}L_{train}
\end{aligned}
\end{equation}

There are cases where SGD can become very slow. The method of momentum is designed to accelerate learning, especially in the face of high curvature, small but consistent gradients, or noisy gradients \cite{Goodfellow2016}. We modify our training (Algorithm \ref{sgdalg}) to include a velocity variable $v$ storing the momentum by calculating exponentially decaying moving average of past gradients.

\begin{algorithm}
   \caption{Stochastic gradient descent with momentum}
   \label{sgdalg}
\begin{algorithmic}[1]
   \State {\bfseries input:} initial $\mathbf{w}_1$, decays $\boldsymbol{\gamma}$, learning rates $\boldsymbol{\eta}$, loss $L_{train}$
   \State initialize $\mathbf{v}_1 = \mathbf{0}$
   \For{$t=1$ {\bfseries to} $T$}
   \State $\mathbf{g}_t = \nabla_\mathbf{w} L_{train}$
   \State $\mathbf{v}_{t+1} = \gamma_t \mathbf{v}_t - (1 - \gamma_t) \mathbf{g}_t$ 
   \State $\mathbf{w}_{t+1} = \mathbf{w}_t + \eta_t \mathbf{v}_t$
   \EndFor
   \State \textbf{output} trained parameters $\mathbf{w}_T$
\end{algorithmic}
\end{algorithm}

where $\gamma_t$ is the momentum decay rate. The training procedure starts with elementary parameters velocity $v_1=0$ and $w_1$ and ends with $v_T$ and $w_T = w_{T-1} + \eta_{T-1}v_{T-1}$. Algorithm \ref{revsgdalg} is then used to calculate the gradients of validation loss with regard to the hyperparameters.

\begin{algorithm}
   \caption{Reverse-mode differentiation of SGD}
   \label{revsgdalg}
\begin{algorithmic}[1]
   \State {\bfseries input:} $\mathbf{w}_T$, $\mathbf{v}_T$, $\boldsymbol{\gamma}$, $\boldsymbol{\eta}$, train loss $L_{train}$, validation loss $L_{valid}$
   \State initialize $d\mathbf{v} = \mathbf{0}$, $d\boldsymbol{\lambda} = \mathbf{0}$, $d\eta_t = \mathbf{0}$, $d\gamma = \mathbf{0}$
   \State initialize $d\mathbf{w} = \nabla_\mathbf{w} L_{valid}$
   \For{$t=T$ {\bfseries counting down to} $1$}
   \State $d\eta_t = d\mathbf{w}^{\mathsf{T}} \mathbf{v}_t$
   \State $\mathbf{w}_{t-1} = \mathbf{w}_t - \eta_t \mathbf{v}_t$ 
   \State $\mathbf{g}_t = \nabla_\mathbf{w} L(\mathbf{w}_t, \boldsymbol{\lambda}, t)$ 
   \State $\mathbf{v}_{t-1} = [\mathbf{v}_t + (1 - \gamma_t) \mathbf{g}_t] / \gamma_t$ 
   \State $d\mathbf{v} = d\mathbf{v} + \eta_t d\mathbf{w}$
   \State $d\gamma_t = d\mathbf{v}^{\mathsf{T}} (\mathbf{v}_t + \mathbf{g}_t)$
   \State $d\mathbf{w} = d\mathbf{w} - (1 - \gamma_t) d\mathbf{v} \nabla_\mathbf{w} \nabla_\mathbf{w} L_{train}$ 
   \State $d\boldsymbol{\lambda} = d\boldsymbol{\lambda} - (1 - \gamma_t) d\mathbf{v} \nabla_{\boldsymbol{\lambda}} \nabla_\mathbf{w} L_{train}$
   \State $d\mathbf{v} = \gamma_t d\mathbf{v}$
   \EndFor
   \State \textbf{output} gradient of $L_{valid}$ w.r.t $\boldsymbol{\lambda}$
\end{algorithmic}
\end{algorithm}

The velocity $v_t$ is needed to reverse the path, otherwise without momentum, $g_t$ and $\eta_t$ alone would not be able to recover $w_{t-1}$. Notice that the loss of information caused by finite precision arithmetic in computers leads to failure of this algorithm. For this reason, we need to store the bits lost in $v_t$ when multiplied by $\gamma_{t}$.

Given this powerful gradient-based mechanism for finding hyperparameters, a natural extension to our model is to introduce a hyperparameter $\alpha$ denoting the contribution of skip-layer and dense-network in producing predictions with higher generalization. That is to say, our model can be reformulated as:
\begin{equation} \label{NN2}
y_t= \Phi({\bm x};{\bm w}) =  \alpha\sum_{i\rightarrow k} x_{it}w_{ik} +(1-\alpha)\sum_{j\rightarrow k} \phi_j  \bigg(\sum_{i\rightarrow j} x_{it} w_{ij} \bigg)w_{jk}+\varepsilon_t,
\end{equation}

where $\alpha$ assumes a value between zero and one. Appreciating that the skip-layer and the dense network have unbalanced effect on the outcome, one can see how this may result in faster convergence of training procedure. More importantly, $\alpha$ can be interpreted as the activation of skip-layer and dense network and can point to linearity or nonlinearity components.

\section{Case Study: Return Prediction}

Research into modelling and forecasting financial returns has a long history. Several models are described in \cite{tsay2005} and \cite{campbell1996} that attempt to explain return time series using linear combinations of one or more financial market factors. The most widely studied single factor model is the capital asset pricing model (CAPM) of \cite{sharpe1964} and \cite{lintner1965} that relates the expected return of equities to the expected rate of return on a market index (such as the Standard and Poor’s 500 Index). The empirical performance of CAPM is poor as it cannot explain the behaviour of asset returns, see \cite{fama2004}. This failure is perhaps due to the absence of multiple factors. 
Arbitrage pricing theory (APT) is a general model proposed by \cite{ross1976} to account for these deficiencies. APT presents a linear approximate model of expected asset returns based on an unknown number of macroeconomic ``factors'' or market indices. The relationship between the factors and historical returns is routinely determined linearly.

Return time series present characteristics such as comovement, nonlinearity, non-Gausianity (skewness and heavy tails), volatility clustering and leverage effect. This makes the modelling task very challenging, see \cite{hsieh1991}; \cite{bollerslev1994}; \cite{brooks1996}; \cite{cont2001}. 

The data are daily returns of $ m =$ 418  equities on the S\&P 500 index from 03.01.2006 through 28.09.2018, for a total of 3208 observations. The initial sample 03.01.2006 - 28.09.2017 is used for estimation (training), with $T =$ 2957 in-sample size. The holdout sample period 01.10.2017 - 28.09.2018 (251 observations) is employed to examine the models' out-of-sample forecasting performance. 1-step (here one day) ahead forecasts of targets ($ \hat y_{it+1|t} $) are based on a rolling estimation window. Parameter estimates are updated every five steps.

We believe accounting for comovements between financial returns is important in forecasting returns. Consequently, the lags of other equities are included as predictors for any return series. We examine the nonlinear high-dimensional forecasting model described in the prior sections (AAShNet model) as well as several competing models and benchmarks. 

We compare our proposed model with a benchmark, the sample mean of ${\bm y}_t$ over the in-sample window, as the 1-step ahead forecast. This corresponds to assuming the log daily price of follows a random walk (RW) with drift. It is almost equivalent to the \enquote{zero forecast} when the in-sample window is large enough. Furthermore, a buy-and-hold (B\&H) strategy in the market portfolio (S\&P 500
Index) has been considered as another benchmark. To understand whether allowing nonlinearity improves portfolio performance we examine the AAShNet algorithm (with Ridge and Lasso) optimized by cross-validation.

Since predictability of financial returns has major consequences for financial decision making, the model with minimal forecast error is deemed optimal. However, the model with minimum forecast error does not necessarily guarantee profit maximization, the primary objective of financial decision makers. \cite{armstrong1992}, \cite{pesaran1995,pesaran2000}, \cite{granger2000} and \cite{engle2006} argue that a forecast evaluation criterion
should be related to decision making and judge predictability of financial returns in terms of portfolio simulation. More specifically, a trading (portfolio) simulation approach assumes that all competing models are applied with stock market virtual investment decisions, and out-of-sample portfolio performances are used
to evaluate the predictability of alternative models.

\begin{figure}[!h]

\centering
\includegraphics[width= 1.05 \columnwidth]{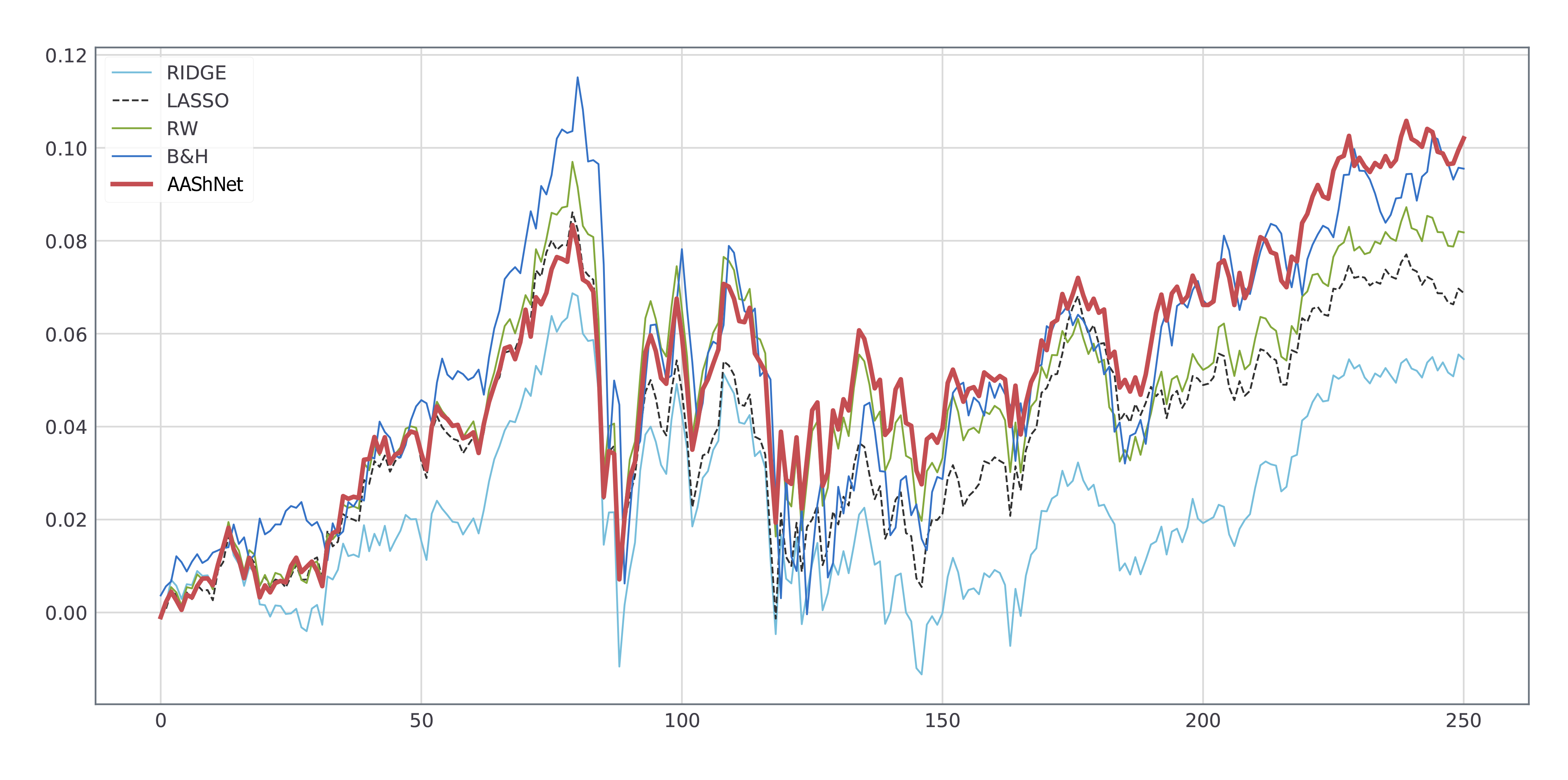}
\caption{Comparison of AAShNet, and the competing models based on the portfolio excess returns in the out-of-sample period.}
\label{fig:results}
\end{figure}

Consequently, this paper examines both statistical and portfolio performance measures (the out-of-sample RMSE and the portfolio performance during the out-of-sample period). Figure \ref{fig:results} illustrates portfolio excess returns for the out-of-sample period for the proposed model (ASShNet) against competing approaches. We randomly selected 50 stocks out of 418 stocks to construct the portfolio. However, the forecast of each selected stock is based on the lags of all 418 equities.

Consider a passive, equally weighted $(1/M)$ portfolios with short selling. This portfolio is known to be a very stringent benchmark that many optimization models fail to outperform (see \cite{demiguel2009}. We compute the portfolio’s out-of-sample excess returns and volatility as well as the Sharpe ratio. Sharpe ratio measures risk-adjusted returns, a portfolio with a greater Sharpe ratio offers greater returns for the same risk. If a portfolio with lower Sharpe ratio has returned better over a time period than another portfolio with a higher ratio, the risk of losing by investing in the former fund will be higher.

\begin{table}[!htbp]
\centering
\caption{}
\begin{tabular}{ | l | r | r | r | r| r|}
    \hline
    Portfolio & Return & Sharp Ratio & Ave(RMSE) \\ \hline \hline
    AAShNet & 10.2\% & 1.143 & 0.01558\\ \hline
    B\&H & 9.55\% & 0.767 & -\\ \hline
    RW & 8.17\% & 0.770 & 0.01564\\ \hline
    Ridge & 5.45\% & 0.561 & 0.01613\\
    \hline
    Lasso & 6.87\% & 0.741 & 0.01590\\
    \hline
   \end{tabular}
    \label{tab:results}
\end{table}

The proposed penalized neural net behaves noticeably better in this empirical analysis. Table \ref{tab:results} provides evidence for out-of-sample forecasting ability of this model vis-\`a-vis  competing approaches in terms of the Sharp ratio. AAShNet also offers an appreciable improvement over linear shrinkage models and benchmarks based on RMSE and actual portfolio performance. In the Ridge and Lasso regressions, the best model is selected by cross-validation. We perform generalized cross-validation, which is an efficient leave-one-out cross-validation.

AAShNet produces higher returns (10.24\%) at the end of the out-of-sample period, with a Sharpe ratio of (1.143) that is superior to alternative models. This indicates that significantly improved forecast is obtained by modelling nonlinear dynamics among variables. One should note that Random Walk with drift and AR(1) are special cases of shrinkage models and AAShNet when there is no dependence on other equities.

\section{Concluding Remarks}

Forecasting with many predictors has received a good deal of attention in recent years.  Shrinkage methods are one of the  most common approaches for forecasting with many predictors. Such methods have generally ignored nonlinear dynamic relations among predictors and the target variable. 

In this study, we suggested an Adaptable and Automated Shrinkage Estimation of Neural Networks (AAShNet). We explained how skip-layer connections move the model in the right direction when the data contains both linear and nonlinear components. To overcome the curse of dimensionality and to manage model complexity, we penalized the model loss function with $L_1$ and $L_2$ norms. Setting the size of shrinkage is still an open question. Recent studies have proposed automated approaches for estimation of algorithm hyperparameters. We employed the gradient-based automated approaches which treat shrinkage hyperparameters in the same manner as the network weights during training, and simultaneously optimize both sets of parameters. 

The empirical application to forecasting daily returns of equities in the S\&P 500 index from 2006 to 2018 provides support for the out-of-sample forecasting ability of AAShNet algorithm vis-\`a-vis some competing approaches, both in terms of statistical criteria and trading simulation performance. Our empirical results encourage further research toward other possible applications of the proposed model.

\section{Appendix: Automatic Differentiation}
There are three main approaches that computer can work out the derivatives: Numerical, Symbolic and Automatic differentiation. 
%Numerical differentiation relies on limit definition of a derivative. Unfortunately its calculation in floating-point arithmetic is inherently ill-conditioned and unstable, and its time complexity would make this approach inappropriate for ML use cases where derivative for millions of parameters are to be calculated. Symbolic differentiation works by symbolic manipulation of closed-form expressions using rules of differential calculus. Symbolic differentiation of an expression tree can be a mechanical process performed by computer that given a formula produces the derivative expressions with respect to independent variables. This approach suffers from the problem of \textit{expression swell}. The final exact form of a derivative of an expression can get exponentially large that it would render the result of symbolic differentiation very difficult to use. {\color{red} we may remove this paragraph if cited works suffice.}
Automatic differentiation refers to a family of procedures to automatically calculate exact derivatives of any function, including program subroutines, with time complexity at most a small constant factor of the time complexity of the original function. It is not inherently ill-conditioned and unstable similar to the numerical method and has much less computational complexity. It also does not suffer from expression swell problem of symbolic differentiation. 

AD augments the standard computation with calculation of derivatives whose combination through chain rule gives the derivative for overall composition. AD can be applied on evaluation trace of arbitrary program subroutines which can be more than closed-form functions and are in fact capable of incorporating complex control flows which do not directly alter values. An automatic differentiator takes a code subroutine that computes a function of several independent variables as input and gives as output a code that computes the original function along the gradient of the function with respect to the independent variables. As most of the functions are piece-wise differentiable and control flows not directly interfering with calculations, chain rule can be used repeatedly in such a way that gradients are calculated along intermediate values being computed. 

Based on modus operandi of automatic differentiation there can be two implementations of this technique; the \textit{forward mode} and the \textit{reverse mode}. We investigate each method, by applying them on the same trivial function $y = f(x_1, x_2) = x_1x_2-cos(x_1)$ at $(x_1, x_2) = (6, 3)$.

\begin{equation}
\begin{aligned}
v_1 &= x_1 &= 6\\
v_2 &= x_2 &= 3\\
v_3 &= v_1v_2 &= 6\times3\\
v_4 &= cos(v_1) &= cos(6)\\
v_5 &= v_3-v_4 &= 18-0.96\\
y &= v_5 &= 17.04
\end{aligned}
\label{eq1}
\end{equation}

In forward mode, we build a \textit{Forward Primal Trace} of the values propagating through the function and a corresponding \textit{Forward Tangent Trace}. Eq. \ref{eq1} shows the forward evaluation of primals. Forward primal trace depicts the natural flow of composition. Eq. \ref{eq2} is the corresponding tangent trace for $\dot{y} = \frac{\partial f}{\partial x_1}$, that is the rate of change of the function $f$ with respect to the input $x_1$. Notice that both traces are evaluated as written, top to bottom. To calculate the derivative with respect to $n$ different parameters, $n$ forward mode differentiations would be needed. This makes the forward-mode very inefficient for deep learning models where the number of parameters may amount to millions.

\begin{equation}
\begin{aligned}
\dot{v_1} &= \dot{x_1} &= 1\\
\dot{v_2} &= \dot{x_2} &= 0\\
\dot{v_3} &= \dot{v_1}v_2+\dot{v_2}v_1 &= 1 \times 3+0 \times 6\\
\dot{v_4} &= \dot{v_1}\times -sin(v_1) &= 1 \times -sin(6)\\
\dot{v_5} &= \dot{v_3}-\dot{v_4} &= 3 - 0.279\\
\dot{y} &= \dot{v_5} &= 2.72
\end{aligned}
\label{eq2}
\end{equation}
 
The reverse mode works by complementing each intermediate variable $v_i$ with an adjoint $\bar{v_i}$ representing the sensitivity of output $y$ to changes in $v_i$. In reverse mode the code is executed and the trace is stored in memory at first stage. At second stage, the adjoints are calculated in opposite direction of the execution of the original function. The reverse adjoint trace corresponding to Eq. \ref{eq1} is depicted in \ref{eq3}.

\begin{equation}
\begin{aligned}
\bar{v_5} &= \bar{y} &= 1 \\
\bar{v_4} &= \bar{v_5}\frac{\partial v_5}{\partial v_4} &= \bar{v_5}\times -1 &= -1 \\
\bar{v_3} &= \bar{v_5}\frac{\partial v_5}{\partial v_3} &= \bar{v_5}\times 1 &= 1 \\
\bar{v_1} &= \bar{v_4}\frac{\partial v_4}{\partial v_1} &= \bar{v_4}\times -sin(v_2) &= -0.27 \\
\bar{v_2} &= \bar{v_3}\frac{\partial v_3}{\partial v_2} &= \bar{v_3}\times v_1  &=  6\\
\bar{v_1} &= \bar{v_1} + \bar{v_3}\frac{\partial v_3}{\partial v_1} &= \bar{v_1} + \bar{v_3} \times v_2 &= 2.72 \\
\bar{x_1} &= \bar{v_1} & &= 2.72\\
\bar{x_2} &= \bar{v_2} & &= 6\\
\end{aligned}
\label{eq3}
\end{equation}

For a function $f:\mathbb{R}^n \rightarrow \mathbb{R}^m$ whose number of operations to be evaluated is denoted by $\textup{ops}(f)$, the complexity of calculating the Jacobian by forward  and reverse modes are $n \times c \times \textup{ops}(f)$ and $m \times c \times \textup{ops}(f)$, respectively, where it is guaranteed that $c < 6$\cite{griewank2008evaluating}. That is if $n\gg m$, backward-mode is preferable, although it would have increased memory requirements. And forward mode should be used when the number of dependent variables is greater than the number of independent variables. 

\clearpage
\end{doublespace}

% Bibliography.
\nocite{*}
%\begin{doublespacing}   % Double-space the bibliography
\bibliographystyle{apalike}
\bibliography{biblio}
%\end{doublespacing}

\clearpage

\end{document}